\documentstyle[12pt]{article}

\parskip=\medskipamount                 
\def\beee{\begin{equation}}
\def\eeee{\end{equation}}
\def\dggg{^{\dagger}}
\def\beq{\begin{eqnarray}}

\def\eeq{\end{eqnarray}}

\begin{document}
\bibliographystyle{unsrt}

\begin{center}
{\Large \bf CANONICAL COMMUTATION RELATIONS IN THE SCHWINGER
MODEL\footnote
{Supported in part by the National Science Foundation, e-mail addresses:
gat@wam.umd.edu, greenberg@umdep1.umd.edu}}\\[7mm]
Gil Gat and O.W. Greenberg\footnote{Supported in
part by a Semester Research Grant
from the University of Maryland, College
Park.}\\
[.1in]
{\it Center for Theoretical Physics\\
Department of Physics and Astronomy\\
University of Maryland\\
College Park, MD~~20742-4111}\\[5mm]
Preprint number 94-087\\[5mm]
{\bf Abstract}
\end{center}

We give the first operator solution of the Schwinger model
that obeys the canonical
commutation relations in a covariant gauge.

{\bf 1. Introduction}

The Schwinger model\cite{sw,ab},
massless two-dimensional quantum electrodynamics,
is an exactly solvable model that exhibits confinement analogous to the
confinement
that is thought to occur in four-dimensional quantum chromodynamics (QCD).
Because of this,
 the model has been extensively discussed in the literature using
both operator and path integral techniques.
Surprisingly, the published operator
solutions do not obey the canonical commutation
relations that follow from the Lagrangian.  The purpose of this letter is to
provide an operator solution that does obey these relations.

{\bf 2.  Review of the Lowenstein-Swieca solution of the massless Schwinger
model}

The Schwinger model is quantum electrodynamics in 1+1 dimensions.  In the
Lorentz gauge,\footnote{This form of the Lagrangian ignores wavefunction
renormalization of the spinor field.  Taking account of this wavefunction
renormalization, the spinor term in the Lagrangian is
$\bar \psi_u i D\!\!\!\! / \psi_u
-\langle \bar \psi_u i D\!\!\!\! / \psi_u \rangle_0=Z\bar \psi i D\!\!\!\! /
\psi$, where $\psi_u$ is the unrenormalized spinor field and
$Z$ is the spinor wavefunction renormalization.  We suppress the
factor $Z$ below.}
\beee
{\cal L}=\bar \psi i D\!\!\!\! / \psi
-\frac{1}{4} F_{\mu \nu} F^{\mu \nu}-\frac{1}{2}(\partial \cdot A)^2,\label{1}
\eeee
where
\beee
D^{\mu}=\partial^{\mu} -e A^{\mu},~~
F^{\mu \nu}=\partial^{\mu}A^{\nu}-\partial^{\nu}A^{\mu}. \label{2}
\eeee

In its massless form this model is exactly solvable. Lowenstein and
Swieca\cite{ls} found
an operator ansatz that yields the matrix elements computed by Schwinger.
Their
ansatz solution has the following form:
\beee
A^{\mu}(x)=-\sqrt{\frac {\pi}{e^2}} (\epsilon^{\mu \nu}\partial_{\nu} \Sigma
+\partial^{\mu} \eta), \label{3}
\eeee
\beee
\psi(x)=:\exp [i \sqrt{\pi} ( \gamma_5 \Sigma(x) - \eta(x))]:\psi^{(0)}(x),
\label{4}
\eeee
where $\eta$ is a free neutral massless field with negative metric
corresponding to the gauge degrees
of freedom satisfying
$[\eta(x),\dot \eta(y)]_{ET}=-i\delta(x^1-y^1)$ , while $\Sigma$ is a free
neutral
massive field with positive metric
representing the physical degrees of freedom satisfying
$[\Sigma(x),\dot \Sigma(y)]_{ET}=i\delta(x^1-y^1)$ and
$\psi^{(0)}$ is a solution of the free massless
Dirac equation.  (ET stands for equal
time.)
This solution displays the main property of the Schwinger model: the only
physical state is a free particle of mass $e /\sqrt{\pi}$.
Lowenstein and Swieca did not discuss the canonical
commutation relations since their interest was to get Schwinger's matrix
elements.

Capri and Ferrari\cite{cf} found another solution of the model,
in covariant
gauges; however, this solution does not obey the canonical commutation
relations.  Both solutions use ansatzes to solve the operator equations; while
this is satisfactory, we believe it can be illuminating to solve the equations
in a more direct way, and a direct method of solution may be helpful in
analyzing the massive model and other models.
The purpose of this letter is to provide a direct solution
that obeys the canonical commutation relations.  We recognize that this is a
rather technical matter; however, in view of the extensive literature on this
model, we think the literature should contain a solution that obeys these
relations.
\newpage
{\bf 3.  Direct operator solution of the Schwinger model}

We use the Lagrangian Eq.(\ref{1}), but drop surface terms so that the gauge
field part becomes
\beee
-\frac{1}{2}(\partial_{\mu}A_{\nu})(\partial^{\mu}A^{\nu}). \label{6}
\eeee
Since the Lorentz group (without inversions) is abelian in 1+1, all irreducible
representations are one-dimensional; thus the vector and spinor fields in the
model are composed of one-dimensional irreducibles arbitrarily pasted together.
We choose to express the Lagrangian in terms of the irreducible fields in the
basis in which
\beee
A^0=\frac{1}{2}(A^++A^-),~~A^1=\frac{1}{2}(A^+-A^-), ~~\psi=(\psi_1,\psi_2),
\label{7}
\eeee
with
\beee
\gamma^0=\left( \begin{array}{cc}
               0&1\\
               1&0
                \end{array} \right),~~
\gamma^1=      \left( \begin{array}{cc}
               0&-1\\
               1&0
                \end{array} \right),~~
\gamma^5=      \left( \begin{array}{cc}
               1&0\\
               0&-1
                \end{array} \right).~~
\label{8}
\eeee
In terms of the irreducible fields,
\beee
{\cal L}=\psi_1\dggg(2i\partial^+ -eA^-)\psi_1+
\psi_2\dggg(2i\partial^- -eA^+)\psi_1+\frac{1}{2}(\partial^1A^+
\partial^1A^- - \partial^0A^+\partial^0A^-). \label{9}
\eeee
where we have introduced lightcone coordinates, $x^+=x^0+x^1,~~x^-=x^0-x^1$.
The corresponding derivatives are
$\frac{\partial}{\partial x^{\pm}}=\frac{1}{2}(\frac{\partial}
{\partial x^0} \pm \frac{\partial}{\partial
x^1})$.  We define these so that $\frac{\partial}{\partial x^{\pm}} x^{\pm}=1$.
Note that although the fields $A^{\pm}$ are lightcone fields, we are {\it not}
using lightcone quantization, but rather are using equal-time canonical
quantization.  The naive operator equations of motion are
\beq
\Box A^+- 2e\psi\dggg_1\psi_1&=&0, \label{10}  \\
\Box A^-- 2e\psi\dggg_2\psi_2&=&0, \label{11} \\
(2i\partial^+-eA^-)\psi_1&=&0, \label{12} \\
(2i\partial^--eA^+)\psi_2&=&0. \label{13}
\eeq
As Schwinger pointed out in his original paper, the spinor bilinear products
require a line integral of the ``vector'' potential
between the $\psi\dggg$ and the $\psi$ in order to ensure gauge invariance;
this
is done explicitly below using point-splitting.
For example, $\psi_1\dggg \psi_1$ is replaced
by
\beee
\lim_{\epsilon \rightarrow 0} \, \frac{1}{2}
\left[
\psi\dggg_1(x+\epsilon)e^{-ie \int^{x+\epsilon}_{x} A_{\mu}(w) dw^{\mu}}
 \psi_1(x)+ cc. \right].  \label{14}
\eeee
The canonical momenta are
\beee
\pi_{A^+}=-\frac{1}{2}\partial^0A^-,~~\pi_{A^-}=-\frac{1}{2}\partial^0A^+,~~
\pi_{\psi_j}=i \psi_j\dggg.
\label{15}
\eeee

First we solve the Dirac equations by exponentiation,
\beee
\psi_1(x)={\cal P}exp[-\frac{ie}{2}\int^{x^+}_{-\infty}A^-(w^+,x^-)dw^+]
\psi_1^{(0)}(x^-),~~\partial^+\psi_1^{(0)}=0, \label{16}
\eeee
\beee
\psi_2(x)={\cal P}exp[-\frac{ie}{2}\int^{x^-}_{-\infty}A^+(x^+,w^-)dw^-]
\psi_2^{(0)}(x^+),~~\partial^-\psi_2^{(0)}=0. \label{17}
\eeee
The point-splitting vector is taken spacelike, $\epsilon=(0,\epsilon^1),
{}~~\epsilon^{\pm}=\pm \epsilon^1$.  Thus, for example, $\psi_1\dggg$ must be
replaced by
\beee
\psi\dggg_1(x+\epsilon)=\psi^{0~\dagger}_1(x^--\epsilon^1)\bar{\cal P}
exp[\frac{ie}{2}\int^{x^++\epsilon^1}_{-\infty}
A^-(w^+,x^--\epsilon^1)dw^+].  \label{18}
\eeee
The symbols ${\cal P}$ and $\bar{\cal P}$ stand for path and antipath ordering,
respectively.
The result of the point-splitting differs from the usual one by having
integrated (nonlocal) terms.
The equations for $A^{\pm}$ become
\beee
(\Box + \frac{e^2}{2 \pi})A^+-\frac{e^2}{2 \pi}\int^{x^+}_{-\infty}
\frac{\partial A^-}{\partial x^-}(w^+,x^-)dw^+=2e\psi^{(0)\dagger}_1(x^-)
\psi^{(0)}_1(x^-) \label{19}
\eeee
\beee
(\Box + \frac{e^2}{2 \pi})A^--\frac{e^2}{2 \pi}\int^{x^-}_{-\infty}
\frac{\partial A^+}{\partial x^+}(x^+,w^-)dw^-=2e\psi^{(0)\dagger}_2(x^+)
\psi^{(0)}_2(x^+). \label{20}
\eeee
The integrated terms here can be removed by taking derivatives with respect
to the upper limit.  Combining the resulting equations leads to
\beee
\Box~~ \partial \cdot A=0 \label{21}
\eeee
\beee
(\Box+\frac{e^2}{\pi})~\epsilon_{\mu \nu}\partial^{\mu}A^{\nu}=0; \label{22}
\eeee
thus $\partial \cdot A \equiv \eta$ is a massless field and
$\epsilon_{\mu \nu}\partial^{\mu}A^{\nu} \equiv \Sigma$ is a field of
mass $e/\sqrt{\pi}$.  As Lowenstein and Swieca observed, $\eta$ and $\Sigma$
(which is the electric field in 1+1) are the gauge-variant and gauge-invariant
degrees of freedom, respectively.
Then $\Box~ A^{\mu}$ must be a linear combination of
$\partial^{\mu} \eta$ and $\epsilon^{\mu \nu} \partial_{\nu} \Sigma$.
$A^{\mu}$ must be the convolution of the
$\bar{\Delta}(x)=-\frac{1}{2}\epsilon(x^0)\Delta(x)$ Green's function with
this linear combination plus terms annihilated by $\Box$.
The convolution of
$\bar{\Delta}(x)$
with $\eta$ does not exist, because, formally, it is $\int d^2y \bar
\Delta(x-y)\eta(y) =\int d^2k exp(-ik \cdot x)\delta(k^2) \tilde \eta(k)/k^2$,
which is ill-defined.  Because of this, a new
field $a$ that obeys $\Box~ a =\eta$ must be introduced.  This was  first
done by Capri and Ferrari.
Thus
\beee
A^{\mu}=c_1\partial^{\mu}a+c_2 \epsilon^{\mu \nu}\partial_{\mu}\Sigma
+c_3 \partial ^{\mu} \eta +c_4 \bar{\psi}^{(0)}\gamma^{\mu}\psi^{(0)}.
\label{23}
\eeee
For the massless case,
\beee
\bar{\psi}^{(0)}\gamma^{\mu}\psi^{(0)}=\partial^{\mu} \phi,~~\Box ~\phi=0,
\label{24}
\eeee
where $\phi$ is a free positive-metric scalar field.
Substitution of this $A^{\mu}$ into the integrodifferential
Eqs.(\ref{19},\ref{20}) fixes
$c_4=-2e/\pi$ and leaves the other constants arbitrary.

We want the equal-time canonical commutation relations (for $A^{\mu}$ they are
not renormalized), that play the role of
boundary
conditions, to determine these constants.  To do this, we must find the
equal-time relations among $a$ and $\eta$ and their canonical conjugates.
In order to determine these
relations, we calculate the two-point functions of the fields $\eta$ and
$a$ and assume they are asymptotic fields with c-number commutators.
The coefficients $c_i,~i=1,2,3$ in $A^{\mu}$ take care of the
absolute normalizations;
we allow a metric factor $\tau=\pm 1$ in the two-point function of $\eta$ and
let the CCR's determine the metrics of $\eta$ and $a$.  We use Klaiber's
regulation\cite{kl} of the massless scalar two-point function,
\beq
\langle 0|\eta(x)\eta(y)|\rangle&=&\frac{\tau}{2 \pi}\int^{\infty}_{\infty}
\frac{dp^1}{2|p^1|}[e^{-ip\cdot \xi}-\theta(\lambda-|p^1|)]  \label{25} \\
&=&-\frac{\tau}{4 \pi}ln(-\mu^2 \xi^2+i\xi^0 \epsilon)  \label{26}\\
&\equiv & D^{(+)}(\xi;\lambda), \label{27}
\eeq
where $~\xi=x-y,~~\mu=e^{\gamma}\lambda,~~\gamma =
{\rm Euler's~~ const.}$
 From $\Box~ \langle 0|a(x)\eta(y)|0\rangle=\langle 0|\eta(x)\eta(y)|0\rangle$,
we find
\beee
\langle 0|a(x)\eta(y)|\rangle=\tau I^{(+)}(\xi;\lambda)+
c_{a \eta}D^{(+)}(\xi;\lambda), \label{28}
\eeee
\beee
I^{(+)}(\xi;\lambda)=-\frac{1}{4 \pi}[\frac{\xi^2}{4}ln(-\mu^2 \xi^2+i\epsilon
\xi^0)-\frac{1}{2}\xi^2], \label{29}
\eeee
where the $D^{(+)}$ term is an arbitrary solution of the homogeneous equation.
(Capri and Ferrari's result for $I^{(+)}$ is incorrect).
Analogously,
\beee
\langle 0|\eta(x)a(y)|\rangle=\tau I^{(+)}(\xi;\lambda)+
c_{\eta a}D^{(+)}(\xi;\lambda). \label{30}
\eeee
The two-point function
of two $a$'s can be calculated in two ways; consistency requires
$c_{a \eta}=c_{\eta a}$.  Then
\beee
\langle 0|a(x)a(y)|\rangle=\tau K^{(+)}(\xi)+c_{a \eta}I^{(+)}(\xi;\lambda)+
c_{aa}D^{(+)}(\xi;\lambda), \label{31}
\eeee
\beee
K^{(+)}(\xi)=-\frac{1}{256 \pi}(\xi^2)^2[ln(-\mu^2 \xi^2+i\epsilon \xi^0)-3],
\label{32}
\eeee
where we introduce a new function, $K^{(+)}$.
The ETCR's follow simply from the assumption that the $a$ and $\eta$ fields
are free.
\beee
[\eta,\dot{\eta}]_{ET}=\tau i \delta,~~[\eta,\dot{a}]_{ET}=c_{a\eta} i
\delta,~~
[a,\dot{\eta}]_{ET}=c_{a\eta} i \delta,~~[a,\dot{a}]_{ET}=c_{aa} i \delta,~~
\label{33}
\eeee
The extra terms that arise from solutions of the homogeneous equations for
the two-point functions allow a lot of freedom.  The system is much more
constrained if these terms are all set to zero, i.e., $c_{a \eta}=c_{aa}=0$;
in this case there are
two solutions that obey the CCR's:
\beee
\tau=-1,~~c_1=\mp \sqrt{\frac{e^2}{5 \pi}},~~c_2=\pm \sqrt{\frac{\pi}{e^2}},~~
c_3=\pm \sqrt{\frac{5 \pi}{e^2}}. \label{34}
\eeee
The sign ambiguity is due to the fact that the CCR's are bilinear in $A^{\mu}$
and $\dot{A}^{\mu}$.
As expected, $\tau=-1$ shows that the fields $\eta$ and $a$ that are associated
with the gauge-variant degree of freedom are ghosts.

{\bf 4. Conclusions}

The canonical commutation relations in field theory and their predecessors in
classical mechanics and quantum mechanics are important for many reasons.
The Poisson brackets in classical mechanics, for example,
ensure
that the Hamiltonian is the generator of time translations.  In
quantum mechanics, for example,
the relation $[x,p]=i\hbar$ leads to the uncertainty relation.
In quantum field
theory, the CCR's lead to the free field being a collection of quantized
oscillators.  In nonrelativistic field theories at least, the CCR's imply
unitarity\cite{cg}.  A new feature of the canonical commutation
relations in quantum field
theory is that they ensure that the asymptotic fields have the proper free
commutation relation.  (The renormalized canonical commutation relations will
do
as well as the original CCR's for this purpose.)
For these reasons, a solution that obeys the canonical commutation
relations is important.

{\bf Acknowledgements}

It is a pleasure to thank Joseph Sucher for reading the manuscript and for
making good suggestions.

\end{document}